\documentclass{article}

\usepackage{arxiv}

\usepackage[utf8]{inputenc} 
\usepackage[T1]{fontenc}    
\usepackage{hyperref}       
\usepackage{url}            
\usepackage{booktabs}       
\usepackage{amsfonts}       
\usepackage{nicefrac}       
\usepackage{microtype}      
\usepackage{lipsum}
\usepackage{graphicx}
\usepackage[T1]{fontenc} 
\usepackage{graphicx,color}
\usepackage[dvipsnames]{xcolor}
\usepackage{soul}
\usepackage{url}
\usepackage{subcaption}
\usepackage{amssymb}
\usepackage{float}
\usepackage{pdf14}
\usepackage{tikz}
\usepackage{placeins}

\usepackage{color}

\newcommand{\changes}[1]{\textcolor{black}{#1}}
\graphicspath{ {./img/} }

\title{Neural Networks Optimized by Genetic Algorithms in Cosmology}

\author{
Isidro G\'omez-Vargas \\
  Instituto de Ciencias Físicas, \\ Universidad Nacional Aut\'onoma de M\'exico,\\
  62210, Cuernavaca, Morelos, M\'exico.\\
  \texttt{igomez@icf.unam.mx} \\
\And
Joshua Briones Andrade\\
  Facultad de Ciencias, \\ Universidad Nacional Aut\'onoma de M\'exico,\\
  Ciudad de M\'exico, M\'exico.\\
  \texttt{joshuabriones@ciencias.unam.mx} \\
\And
  J. Alberto Vázquez\\
  Instituto de Ciencias Físicas, \\ Universidad Nacional Aut\'onoma de M\'exico,\\
  62210, Cuernavaca, Morelos, M\'exico.\\
  \texttt{javazquez@icf.unam.mx} 
  }

\begin{document}
\maketitle
\begin{abstract}
The applications of artificial neural networks in the cosmological field have shone successfully during the past decade, this is due to their great ability of modeling large amounts of datasets and complex nonlinear functions. However, in some cases, their use still remains controversial because their ease of producing inaccurate results when the hyperparameters are not carefully selected. In this paper, to find the optimal combination of hyperparameters to artificial neural networks, we propose to take advantage of the genetic algorithms. As a proof of the concept, we analyze three different cosmological cases to test the performance of the architectures achieved with the genetic algorithms and compare them with the standard process, consisting of a grid with all possible configurations. First, we carry out a model-independent reconstruction of the distance modulus using a type Ia supernovae compilation. Second, the neural networks learn to infer the equation of state for the quintessence model, and finally with the data from a combined redshift catalog the neural networks predict the photometric redshift given six photometric bands (urgizy). We found that the genetic algorithms improve considerably the generation of the neural network architectures, which can ensure more confidence in their physical results because of the better performance in the metrics with respect to the grid method.


\end{abstract}


%
\section{Introduction}
\label{sec:intro}
Throughout cosmology there exists a variety of numerical and statistical techniques that allow to study complex theoretical models and to process large amounts of observational measurements. However, despite the increasing amount of observational evidence, there are still several tensions with unknown physical explanations which encourages the search for new analysis tools to extract valuable information from the data, and to use the computational resources more efficiently. During the last decades, Machine Learning incorporated effective alternatives into the data analysis field, in particular  the Deep Learning. For instance, the Artificial Neural Networks (ANNs) have been successfully used to carry out a broad diversity of different tasks such as regression, classification, image processing and time series, among many others \cite{goodfellow2016deep}. In cosmology, they have achieved great results in performing model-independent reconstructions (nonlinear regression) \cite{escamilla2020deep, wang2020reconstructing, gomez2021neuralarxiv, lin2019non, dialektopoulos2022neural}, in the process of speeding up numerical calculations \cite{graff2012, hortua2020accelerating, spurio2022cosmopower, gomez2021neural, rojas2022observational} or in the classification of different objects \cite{perraudin2019deepsphere, ishida2019machine, moller2020supernnova, rojas2022observational}. Nevertheless, in most of these studies the architecture of the network is built up by generating a grid with a large number of possible combinations of hyperparameter values in order to select the best one, which could be computationally expensive; and in other scenarios the architecture is constructed on an empirical way, which may lead to inaccurate results.

Despite their great advantages, ANNs have two main drawbacks. First, the fact that artificial networks have thousands or millions of parameters (called weights) generates a hard interpretation of them. Second, ANNs have also several hyperparameters that must be picked out carefully (e.g. number of layers, nodes, activation function, batch size, etc.) in order to have acceptable predictions and therefore the results depend on their selection. That is, even though several combinations are able to generate a good neural network model, there are even more bad combinations that will achieve incorrect predictions and, in cosmology, a spurious and imprecise physical interpretation. If the hyperparameters configuration is adequate, there is a good balance between the bias and variance of the neural model which implies the network is neither over-fitted nor under-fitted \cite{geman1992neural}, therefore the predictions should be reliable and thus underestimate the weak interpretation of the multiple weights. 

There are several strategies for finding the appropriate values of the hyperparameters in a neural network \cite{larochelle2007empirical, hutter2009paramils, bardenet2013collaborative, zhang2019deep}. The standard approach is to propose a multidimensional grid form of several values of the hyperparameters \cite{larochelle2007empirical}, evaluate all possible combinations and determine, by comparison, which combination has the best performance. Newer approaches are based in mathematical optimization or metaheuristic algorithms, both of which consist in specialized algorithms to find the optimal value for a given function. Therefore, the search for the best combination of hyperparameters of neural networks is posed as an optimization problem. 
Genetic algorithms, by themselves, have been investigated in cosmology for quasar parameterisation \cite{wasiewicz2019optimization}, nonparametric reconstructions \cite{arjona2022testing, arjona2020machine}, string theory analysis \cite{cole2019searching}, to name just a few. In other research areas, there are various cases in which artificial neural networks and genetic algorithms have been applied together; for example, in medicine \cite{drugdiscovery, braintumor, anaraki2019magnetic}, seismology \cite{seismicsignals} and in string theory \cite{evolvingnn}, among others.

In several fields it is still  uncommon to pay particular attention to the hyperparameter tuning during the construction of the neural network model. Regularly, an usual strategy is the hyperparameter grid and in several cases may be a good enough option, however there are better ways to optimize this selection. In this paper, we explore the use of the Genetic Algorithms (GA), the most popular metaheuristic algorithm, and compare their performance with the standard grid of hyperparameters. In this context, the goal of the genetic algorithm is to find the best combination of neural network hyperparameters that minimizes the target function.

This paper is structured as follows. In the next section we provide a brief overview about neural networks, genetic algorithms and the hyperparameter tuning approaches. In Section \ref{sec:methodoloy} we explain the technical details about our implementation. Section \ref{sec:results}, as a proof of the concept, contains three cosmological study cases, in the first one we made a model-independent reconstruction of the distance modulus using a Type Ia Supernovae compilation; in the second we train neural network models using the analytical results of \changes{Equation of State of the Quintessence model}; and last, a \changes{prediction of the photometric redshift given some features from the catalog of Reference \cite{zhou2019deep}}. In Section \ref{sec:conclusions} we describe our final remarks about this research. \changes{In addition, in Appendix \ref{appendixA} we consider some randomness effects in the case study of Section \ref{sec:case1_rec} to verify the robustness of the genetic algorithms in the context of this work.}
 
\section{Machine learning background}
\label{sec:background}
Machine learning is the field of Artificial Intelligence focused on the mathematical modeling of the data, it extracts the intrinsic properties of datasets by minimizing an objective function through many iterations until an acceptable combination of model parameters is found. In recent years, the most successful types of machine learning models are artificial neural networks, which have thousands or millions of parameters, called weights, that allow modeling any nonlinear function \cite{hornik1990universal}. 
Finding the correct combination of hyperparameters, or in other words the best neural network architecture, is a hard task. The classic method for this is to generate several combinations of hyperparameters and evaluate all of them until the best one is found. In recent years, by taking advantage of the existing computing power, various optimization and parameter estimation techniques have been applied to find these hyperparameters more efficiently. In particular, the metaheuristic optimization algorithms (i.e. the genetic algorithm), which allow finding the best solution to an optimization problem without using derivatives.

In this section we describe very briefly artificial neural networks, genetic algorithms and the hyperparameter tuning.

\subsection{Artificial neural networks}    
\label{sec:ann}
Artificial neural networks have been applied in various scientific fields because of their ability to model large and complex datasets. The universal approximation theorem guarantees that ANNs can model any nonlinear function \cite{hornik1990universal}, making them a powerful tool in modeling datasets where the intrinsic relationships of their variables are unknown and most of the time multidimensional. A complete review of neural networks is beyond the scope of this article; there are great references in the literature to delve into this topic in a formal way \cite{goodfellow2016deep, bishop2006pattern, nielsen2015neural}; for a basic introduction, their main algorithms and a cosmological context, see \cite{rojas2022observational}.

Inspired by nature, an artificial neural network (ANN) consists of a computational model that aims to emulate the synapse through interconnected layers of units called neurons or nodes, which make up its basic information processing elements. In the simplest type of network, the feedforward neural network, there are three types of layers: an input layer that receives the initial information, hidden layers responsible for extracting patterns and producing nonlinearity effects, and finally the output layer that presents the results of the prediction.

The intrinsic parameters of ANNs are known as hyperparameters, which, unlike the weights, are not adjusted during the training and must be configured in advanced. Examples of hyperparameters are the number of layers, the number of nodes per layer, the number of epochs and the activation function. In addition, since ANNs use a gradient descent and a backpropagation algorithm \cite{rumelhart1986learning}, the parameters of these algorithms can also be hyperparameters of the ANN model, e.g., batch size and learning rate, among others. In practice, some hyperparameters are fixed and others remain as free parameters that are found by a tuning strategy. 

\subsection{Genetic algorithms}    
    \label{sec:ga}

Genetic algorithms are inspired by genetic populations which consider any possible solution of an optimization problem as an individual\changes{, or chromosome}. They are popular for their ability to solve large-scale nonlinear and nonconvex optimization problems \cite{gallagher1994genetic} and for difficult search situations \cite{sivanandam2008genetic}. It is beyond the scope of this article a full background of genetic algorithms, but for interested readers we recommend the following references \cite{reeves1997genetic, katoch2021review, wirsansky2020hands}. 

Inspired by genetic populations, the first step of the algorithm is to generate several individuals, within the search space, and define them as a population. Then, through operations such as crossover, mutation and selection over several iterations (called generations) the population gets closer to the optimum of a target function. In any problem approached with genetic algorithms, it is necessary to select the fitness, or target function to optimize, to define the search space and to set the genetic operators (selection, crossover and mutation).

\changes{Selection operation is the criterion used to determine which individuals, at given generation, will survive or reproduce in the next generation. There are several types of selection methods such as  roulette wheel, stochastic universal sampling, ranking or tournament. In this paper, we use the tournament selection, for which, in each round two or more individuals, randomly selected, are confronted and the one with the highest fitness, or target function, wins and survives.} 

\changes{The crossover operation, also called recombination, interchanges the genes of two individuals generating an offspring, i.e., a new individual. Usually, the probability of crossover is high; but zero means that the two parents pass to the next generation without doing anything else. There are several crossover types, however in this work we focused in the uniform crossover: for each gene of the new individual it chooses randomly one of the parents.} 

\changes{On the other hand, mutation refers to a random change in one or more genes values in an individual. We use the bit flip mutation \cite{chicano2015fitness}, that consists in the random selection of one gene to be flipped, i.e., to be switched to different value, for example 0 to 1 or vice versa. We need to avoid high values for the mutation probability, because it can become a random walk instead of an effective exploration of the search space.}

At the beginning, it is necessary to assign probability values to the crossover, to the mutation operators and for each iteration two individuals can have a crossover or a single individual a mutation with these probabilities. The value of elitism indicates how many individuals are bound to pass to the next generation, so it is a positive integer value. Thus, in a few words the genetic algorithm works as follow: it generates an initial population within the search space and, generation by generation, the individuals are modified by the operators and by evaluating the objective function, the individuals are approaching the optimum of the target function.

\subsection{Neural Networks hyperparameter tuning}
To find a good combination of the neural network hyperparameters, we focus on two approaches: the classic grid of hyperparameters and the genetic algorithms. Here are some highlights of both.

\subsubsection{Conventional grid}

The typical approach to finding a correct combination of hyperparameters in the ANN is to go through an array of possible hyperparameter values and evaluate each combination to choose the best that minimizes the neural network loss function \cite{larochelle2007empirical}. This involves training as many times as there are combinations, and it is very computationally expensive. Another technique that attempts to reduce this cost is a random search, in which hyperparameter combinations are randomly sampled; however, it still has the same problems and depends on the size of the search space for its efficiency. In both approaches the best solution is always within the initial set of combinations and, however, a lot of configurations have to be evaluated. 

\subsubsection{Using genetic algorithms}

The search for the hyperparameters of an ANN can be considered an optimization problem. Due to the increased number of hyperparameters, the search space is likely to be complex and high-dimensional. Classical optimization methods involving derivatives can be very difficult to implement in this kind of scenario, therefore genetic algorithms are a very interesting way to solve this problem.

The crucial step in using genetic algorithms in hyperparameter fitting is to define the fitness function, or target function. For the case of neural networks, the loss function can be used as a fitness. The loss function during neural network training aims to be minimized, therefore the task of genetic algorithms is to find the best combination of hyperparameters that minimizes the target function. Several research works, unrelated to cosmology, have already combined these two powerful tools and in most cases have promising results \cite{miller1989designing, montana1989training, aszemi2019hyperparameter,  ma2018road}.


\section{Methodology}
\label{sec:methodoloy}
We use \texttt{tensorflow} \cite{abadi2016tensorflow} to program the ANN models and the \texttt{DEAP} library \cite{DEAP_JMLR2012, de2012deap, de2014deap} to implement the genetic algorithms, both in \texttt{Python}. We developed a \texttt{Python} library called \texttt{NNOGADA}\footnote{Repository for NNOGADA (Neural Networks Optimized by Genetic Algorithms for Data Analysis) available at: \url{https://github.com/igomezv/nnogada}} in which a simple genetic algorithm searches the best hyperparameters for a neural network. 

In this framework, the target function of the genetic algorithm must be some metric of the Artificial Neural Network; typically it would be the loss function, but for classification problems it could be accuracy, precision or something similar. In the case of the loss function, the problem would be a minimization and in the case of a classification metric, a maximization. Throughout all cases in our study, we use the loss function and therefore have minimization problems.

The first step is to define the hyperparameters of the neural network model to be found. In practice, there are some hyperparameters that have values recommended by the literature or by experienced users on certain types of problems. In this work, as variable hyperparameters we choose the number of layers, the number of nodes, the learning rate and the batch size. With these we define the search space of the optimization problem. The gradient descent algorithm in our neural network models is \texttt{Adam} \cite{kingma2014adam}, and its hyperparameter that we tune is precisely the learning rate. Secondly, it is necessary to define the possible values for each hyperparameter, where the hyperparameter grid algorithm will search for the best combination, and those that the genetic algorithm will use to generate the first population. 

In the case of the genetic algorithm, the possible values of the free hyperparameters must be encoded in a way that the genetic algorithm can understand them (binary, hexadecimal, etc.). Next, it is necessary to define the population size and the number of generations. Also the probability of mutation, crossover and elitism. Once this configuration of the genetic algorithm is established, we can use the neural model as a function to optimize. To have a fair comparison, we chose these parameters of the genetic algorithms to have a number of neural network evaluations similar to the hyperparameter grid cases; the formal selection of the parameters of the genetic algorithm is out of the scope of this paper.

\begin{figure*}[t!]
    \centering
    \captionsetup{justification=raggedright, singlelinecheck=false, font=footnotesize} 
    \includegraphics[trim=80mm 65mm 80mm 65mm, clip, width=13cm, height=6.5cm]{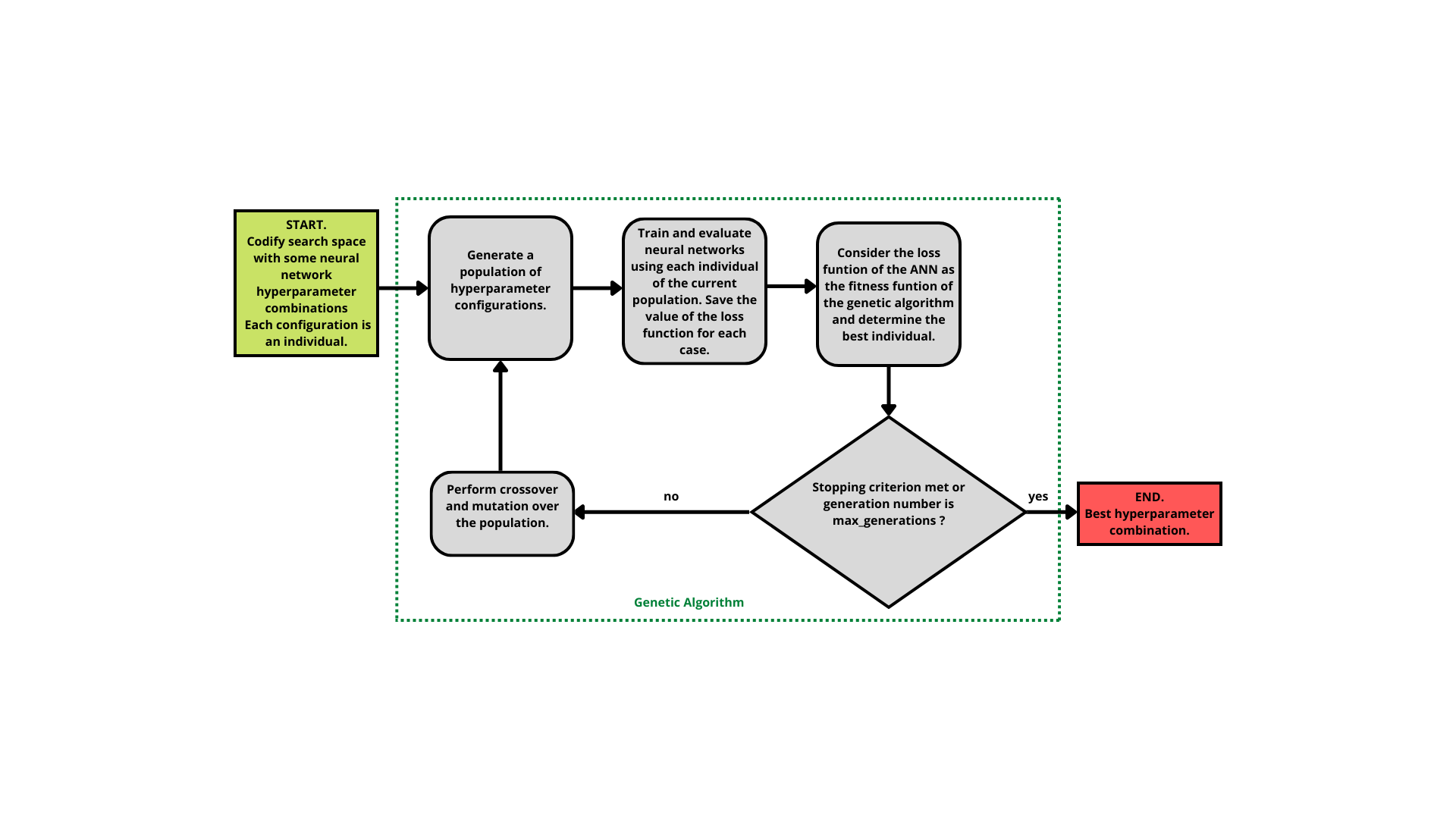}
    \caption{Diagram of neural network hyperparameter tuning with a genetic algorithm.}
    \label{fig:genetic_tuning}
\end{figure*}

For the genetic algorithms, in all of the following case studies in Section \ref{sec:results}, we set the tournament method \cite{eremeev2012genetic} for selection with size 2, binary coding, using elitism with a hall of fame size equal to 1. We use the uniform crossover operator and bit flip mutation operator. We varied and tested different values of crossover and mutation probabilities, population size and number of generations with the goal to analyze their effect in the performance of the genetic algorithms. Figure \ref{fig:genetic_tuning} summarizes the implementation of the hyperparameter tuning with genetic algorithms.

In all cases, we split the datasets into a training, validation and test sets. The first is used to adjust the weights during training, the validation set is used to evaluate the performance of each neural network model during training and, finally, the test set contains information not used in training but useful to measure the generalization capability of the neural networks. For the neural networks of the same example, we fix the number of epochs and report the lowest loss function for each case.

\changes{Once the neural networks were trained, to evaluate the performance of the models, we use the Mean Squared Error (MSE), a measurement of the difference between the neural networks prediction and the real values. It is defined as follows: }
\begin{equation}
        {\rm MSE} = \frac{1}{n} \sum_i^n (Y_i - \hat{Y}_i)^2,
\end{equation}
\changes{where $Y_i$ is a vector with predictions of the ANN, $\hat{Y}_i$ a vector with the expected values, and $n$ is the number of predictions (or the length of $Y_i$ and $\hat{Y}_i$).}
\changes{Using the same notation, we also use the Mean Absolute Error (MAE) as a metric}:
\begin{equation}
        \frac{1}{n} \sum_i^n |(Y_i - \hat{Y}_i)|.
\end{equation}

\changes{The last metric we use for our regression cases of study, is the coefficient of determination, also called \textbf{$R^2$}, that provides information about the goodness of fit of a model and it is defined as:} 
    \begin{equation}
        R^2 = 1-\frac{\sum_i^n (Y_i - \hat{Y}_i)^2}{\sum_i^n (Y_i - \overline{Y})^2}, 
    \end{equation}
\changes{where $\overline{Y}$ is the mean of the observed data.}

\section{Cases of study}
\label{sec:results}

We have chosen three different cosmological problems to test the hyperparameter tuning with genetic algorithms and in all of them we use the type of neural networks known as feedforward networks (also called multilayer perceptrons); however, the method can be easily implemented in other cosmological scenarios and with more complex neural network architectures such as convolutional or recurrent. In the following examples, we compare the performance of a hyperparameter grid and the genetic algorithms for finding an acceptable combination of hyperparameters in neural network models. We test these two strategies in three different cosmological contexts, similar to the problems presented in \cite{rojas2022observational}. The Table \ref{tab:hyp_all} shows our choice of possible hyperparameter values for each of the following subsections; we use these hyperparameters to  construct the grids and the initial populations for the genetic algorithms.

\begin{table}[t!]
    \small
    \centering
    \captionsetup{justification=centering, singlelinecheck=false, font=footnotesize}
    \begin{tabular}{c c c c}
        \hline
         & Section \ref{sec:case1_rec}  & Section \ref{sec:case2_ecs} & Section \ref{sec:case3_photoz} \\
        \hline
        \hline

        Number of layers &    [1, 2, 3, 4] & [1, 2, 3, 4] & [3, 4] \\
        \hline
        Number of nodes & [50, 100, 150, 200] & [50, 100, 150, 200] & [100, 200] \\
        \hline
        \vspace{0.1cm}
       Learning rate  & [$10^{-4}$,$10^{-3}$] & [$10^{-4}$,$10^{-3}$] & [$10^{-4}$,$10^{-3}$]\\
       \hline
       Batch size & [2, 4, 8, 16] & [8, 16] & [8, 16, 32, 64] \\
       \hline
       \hline
    \end{tabular}
    \caption{Hyperparameters for the three cases of the Section \ref{sec:results}.}
    \label{tab:hyp_all}
\end{table}

\subsection{Reconstruction of distance modulus} 
\label{sec:case1_rec}

Model-independent cosmological reconstructions refer to the use of statistical or computational techniques to generate a model of a cosmological observable without assuming an underlying theory. In cosmology, several techniques have been used such as  histogram density estimators \cite{sahni2006reconstructing}, Principal Component Analysis (PCA) \cite{sharma2020reconstruction, escamilla2021model}, smoothed step functions \cite{gerardi2019reconstruction}, gaussian processes \cite{williams2006gaussian, Keeley:2020aym, l2020defying, mukherjee2022revisiting}, extrapolation methods \cite{montiel2014nonparametric}, Bayesian nodal free-form methods \cite{Vazquez:2012ce, hee2017constraining}, evolutionary algorithms \cite{nesseris2012new, arjona2020machine, arjona2022testing} and recently, neural networks \cite{escamilla2020deep, wang2020reconstructing, dialektopoulos2022neural, gomez2021neuralarxiv}.

In this example, we perform a reconstruction of the distance modulus \changes{$\mu(z)$, using data from the Joint Lightcurve Analysis (JLA), with 740 Type Ia supernovae \cite{betoule2014improved}}, that already have been approached with this type of computational models \cite{escamilla2020deep, wang2020reconstructing, gomez2021neuralarxiv}, sometimes with a grid of hyperparameters for tuning the neural network architecture and others without any criterion.
We assume a spatially flat universe, for which the relationship between  the luminosity distance $d_L$ and  the comoving distance $D(z)$ is given by: 
      \begin{equation}
         d_L (z) = \frac{1}{H_0}(1+z)D(z), \qquad {\rm with }\qquad D(z) = H_0\int \frac{dz}{H(z)}.
      \end{equation}
Thus, the observable quantity is computed by the distance modulus $\mu(z) = 5 \log d_L(z) + 25$.
 \noindent 
 
 The \changes{JLA SNeIa compilation} attributes are: the redshift of the measurement, the distance modulus and its statistical error; in addition, it has a covariance matrix with the systematic errors. We employ the diagonal of the covariance matrix, and with associate errors we add them to the statistical error. Then our neural network should have only one node in the input layer, corresponding to the redshift $z$ and two nodes in the output layer, for the distance modulus and the error (statistical plus systematic).
 
The hyperparameters have been searched \changes{ by training several architectures using the SNIa from the JLA compilation}, varying the number of layers, number of nodes, batch size and learning rate, for the values shown in the second column of the Table \ref{tab:hyp_all}. Then, the grid of hyperparameters evaluates $128$ different neural networks architectures to determine the best. For the genetic algorithms, as it can be seen in the Table \ref{tab:results_jla}, \changes{Case A, B and C} are configurations with different values for the mutation probability, crossover probability, population size and number of generations.

\begin{figure*}[t!]
        \captionsetup{justification=raggedright, singlelinecheck=false, font=footnotesize}        
        \centering
        \makebox[12cm][c]{
        \includegraphics[trim=0mm 0mm 15mm 0mm, clip, width=8cm, height=7cm]{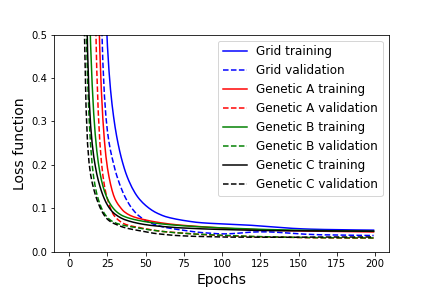}

      \includegraphics[trim=0mm 3mm 15mm 0mm, clip, width=10cm, height=7cm]{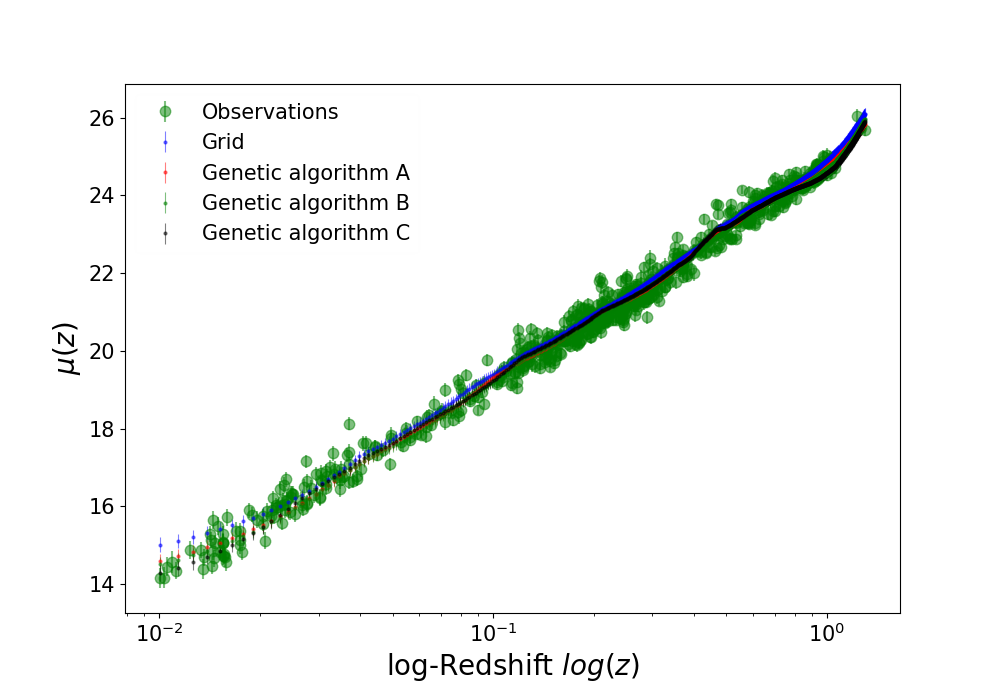}
        }
        \caption{\textit{Left}: Loss function behavior for the training of the neural networks using the JLA dataset. \textit{Right}: Distance modulus reconstruction.}
        \label{fig:jla}
\end{figure*}

In all cases the models were trained along $200$ epochs\changes{, considering the ReLU activation function in all the hidden layers and the linear function in the last one}. In the Table \ref{tab:results_jla} it can be noticed the result for the hyperparameter tuning using a grid and the genetic algorithms \changes{(cases A, B and C)}. For the test set, we use the mean squared error (MSE) \changes{, the mean absolute error (MAE) and the $R^2$ test} to evaluate the performance of the neural network models. We can notice an improvement on the performance for the results of the genetic algorithms, even when it required less evaluations of neural network architectures. Because in this problem the numerical precision is relevant, we can see in Figure \ref{fig:jla} that indeed the genetic algorithms obtain better reconstructions for the distance modulus, particularly in the lower redshifts values. \changes{In the Appendix \ref{appendixA} we validate these results through more repetitions of the algorithms.}

\begin{table}[b!]
    \small
    \centering
    \captionsetup{justification=centering, singlelinecheck=false, font=footnotesize}
    \begin{tabular}{c c c c c}
        \hline
        \hline
          & Grid & Genetic A & Genetic B & Genetic C\\
        \hline
        \hline
        Population size & - & 5 & 5 & 5 \\
        Max generations & - & 10 & 10 & 15 \\
        Crossover & - & 0.5 & 0.5 & 0.5 \\
        Mutation & - & 0.5 & 0.2 &  0.4 \\
      \hline
      \hline
         &  & Hyperparameter results & &  \\
      \hline
      Layers & 4 & 2& 4 & 4 \\
      Nodes & 200 & 100& 100 & 000\\
      Learning rate & 0.0001 & 0.0001 & 0.0001 & 0.0001 \\
      Batch size & 16 & 2 & 4 & 4 \\
      \hline
      \hline
         &  & \changes{Metrics} & &  \\
      \hline
      MSE & 0.0371 & 0.0311 & 0.0314 & 0.0336\\
      \changes{MAE} & 0.1165 & 0.09978 & 0.0985 & 0.1059 \\
      \changes{$R^2$} & 0.4968 & 0.6735 & 0.6797 & 0.7251\\
      \hline
      \hline
      Evaluations & 128 & 41 & 34 & 50\\
      
      \hline
      \hline
    \end{tabular}
    \caption{Results of neural nets training with JLA compilation.}
    \label{tab:results_jla}
\end{table}

Our results suggest that, in previous cosmological works in which the reconstructions are performed with neural networks, there is a possibility that a better architecture could be found using genetic algorithms instead of the use of the hyperparameter grid, and evidently for cases where no strategy is employed to find the correct architecture. 

\subsection{\changes{Quintessence Equation of State}}
\label{sec:case2_ecs}
In this case, we test the capability of neural networks to modeling the Equation of State (EoS) of an scalar field $\phi$ for the Quintessence model, which is computed from a set of differential equations. This is a common and non trivial problem because it involves a shooting method to find the right initial conditions
(see an example in \cite{vazquez2021bayesian}). In particular, the EoS of the Quintessence model is defined as follows:
\begin{equation}
w(z) = \frac{\dot{\phi}^2 - 2V( \phi)}{  \dot{\phi}^2 + 2V( \phi)} ,
\end{equation}
with a potential $V$ and the derivative of the field $\dot{\phi}$, that, in a Friedman-Robertson-Walker background, satisfy the Klein-Gordon equation:
\begin{equation}
\ddot\phi + 3H\dot\phi +\frac{\partial V( \phi)}{\partial\phi} = 0,
\end{equation}
where $H$ is the Hubble parameter
\begin{equation}
    H^2 = \frac{1}{3}\left[\frac{1}{2}\dot{\phi}^2 + V( \phi) + \rho_{M} \right].
\end{equation}

For simplicity, we use the scalar field potential $
    V =  \frac{1}{2}m_{\phi}^2\phi^2,$ with $m_{\phi}$ as the mass of the field in units of [3$H_0$], and $\rho_M$ being the matter content (baryons, dark matter) that satisfies the continiuty equation.

We compute the EoS for several redshifts, fixing $H_0=68.2$ and $\Omega_m=0.3$ and varying values for the mass of the field $m_{\phi}$, to generate a dataset of 2800 combinations. We use a modified version of the SimpleMC code \cite{vazquezsimplemc, aubourg2015} to calculate the EoS, including the method to search the initial conditions of the dynamical system \cite{vazquez2021bayesian}. Therefore, we train the neural networks to predict the EoS of the Quintessence model given the redshift and the mass of the field.

Varying the hyperparameters shown in Table \ref{tab:hyp_all} (batch size, learning rate, number of layers and number of nodes), the combinations found with the grid method and the genetic algorithms are included in Table \ref{tab:results_ecs}. 

\begin{table}[t!]
     \small
    \centering
    \captionsetup{justification=centering, singlelinecheck=false, font=footnotesize}
    \begin{tabular}{c c c }
        \hline
        \hline
          & Grid & Genetic\\
        \hline
        \hline
        Population size & - & 8  \\
        Max generations & - & 5  \\
        Crossover & - & 0.8  \\
        Mutation & - & 0.2  \\
      \hline
      \hline
         &  Hyperparameter results &   \\
      \hline
      Layers & 4 & 4  \\
      Nodes & 200 & 200\\
      Learning rate & 0.001 & 0.001 \\
      Batch size & 16 & 8  \\
      \hline
      \hline
         & \changes{Metrics} &    \\
      \hline
      MSE & $0.0008$ & $1.0953\times10^{-5}$ \\
      MAE & 0.0174 & 0.0020 \\
      $R^2$ & 0.9873 & 0.9998 \\
      \hline
      \hline
      Evaluations & 64 & 34 \\
      
      \hline
      \hline
    \end{tabular}
    \caption{Results of ANN training for the Quintessence EoS.}
    \label{tab:results_ecs}
\end{table}

\begin{figure*}[t!]
        \captionsetup{justification=raggedright, singlelinecheck=false, font=footnotesize}
        \centering
        \makebox[12cm][c]{
                \includegraphics[trim= 0mm 0mm 15mm 0mm, clip, width=8cm, height=7cm]{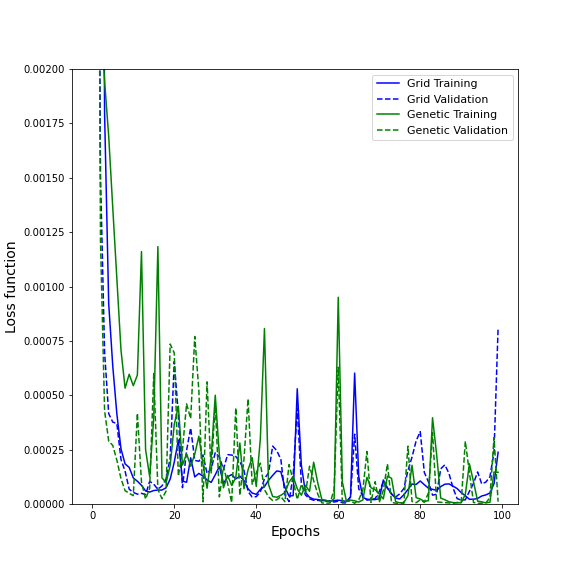}
                \includegraphics[trim= 5mm 0mm 0mm 0mm, clip, width=11cm, height=7cm]{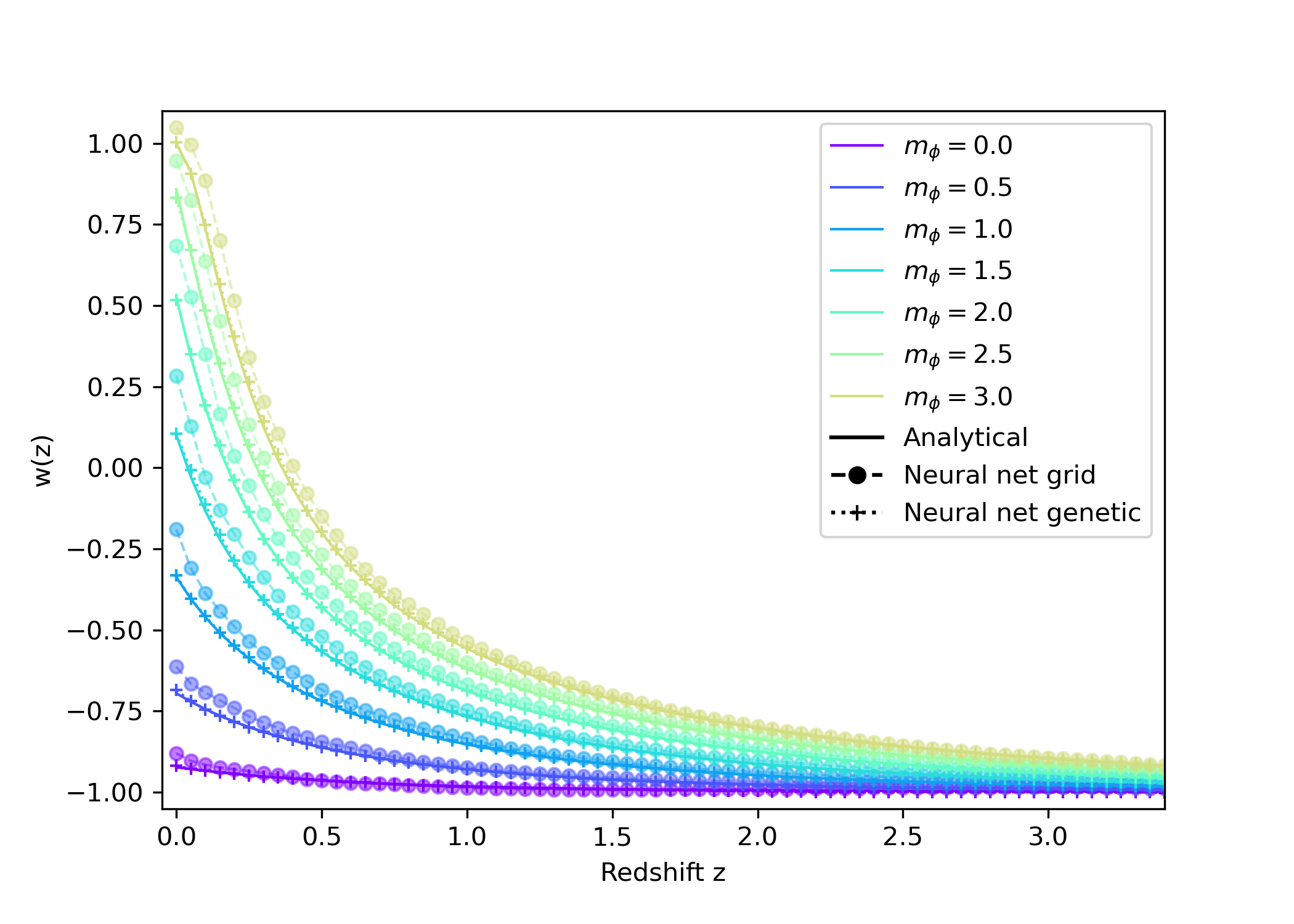}
        }
        \caption{\textit{Left}: Loss function behavior for the the neural networks learning the Equation of State for the Quintessence model. \textit{Right}: Comparison of the analytical EoS for the Quintessence model using different mass values and the predictions with the neural networks.}
        \label{fig:results_eqs}
\end{figure*}

All neural network models were trained with $100$ epochs and using the ReLU activation function for all the hidden layers, and a linear function for the last layer. In Table \ref{tab:results_ecs}, we can see the results of the hyperparameter grid and a genetic algorithm. We can notice that the neural network obtained from the genetic algorithm search has a better performance in the MSE, MAE and $R^2$ metrics, in the case of MSE and MAE it improves with  an order of magnitude, which is very significant. In addition, the genetic algorithms only used 34 neural networks to find the best, while the grid method evaluated 64 different architectures.

In Figure \ref{fig:results_eqs}, it can be noticed the behavior of the loss function for both methods and the ANN predictions of the EoS in comparison with the theoretical values. It is noticeable that the neural network based on the genetic algorithm performs better and their results are closer to the analytical predictions.

\subsection{\changes{Photometric redshift prediction}}
\label{sec:case3_photoz}

In Reference \cite{zhou2019deep}, the authors use  a Random Forest regression to test a catalog that combines photometry from Canada-France-Hawaii Telescope Legacy Survey (CFHTLS), Y-band photometry from the Subaru Suprime camera, and spectroscopic redshifts from DEEP2, DEEP3 \cite{cooper2012deep3} and 3D-HST surveys \cite{brammer20123d}. In this study case, we use the same catalog, see details therein, to test the analysis with neural networks instead of random forest. We generate the six same variables used in \cite{zhou2019deep} based on u,g,r,i,z and Y (see the Table \ref{tab:photoz}). Therefore, the input of the neural networks are these six variables and the output is the photometric redshift.

\begin{table}[t!]
     \small
    \centering
    \captionsetup{justification=centering, singlelinecheck=false, font=footnotesize}
    \begin{tabular}{c c}
    \hline
        Attribute & Description  \\
        \hline
        \hline

        \textbf{u, g, r, i, z} & UV, green, red, near-infrared, far-infrared filters from CFHTLS \cite{hudelot2012vizier}. \\
    \hline
        \textbf{Y} &  Y-band photometry from the SuprimeCam at the Subaru telescope. \cite{miyazaki2002subaru}\\ 
    \hline
    \textbf{redshift} & Photometric redshift.\\
    \hline
    \hline
    \end{tabular}
    \caption{Attributes of the combined catalog of Reference \cite{zhou2019deep}.}
    \label{tab:photoz}
\end{table}

To find the best architecture of the neutwork we use a grid of hyperparameters and genetic algorithms with the combinations shown in Table \ref{tab:hyp_all}. In this case, we vary three hyperparameters: number of layers, number of nodes by layer (the same for all the layers) and the learning rate, such as shown in Table \ref{tab:hyp_all}. The number of epochs is $200$. The loss function fixed is the mean squared error, rectifier linear unit (ReLU) as  an activation function in the hidden layers and the linear function in the last layer.

\begin{table}[t!]
     \small
    \centering
    \captionsetup{justification=centering, singlelinecheck=false, font=footnotesize}
    \begin{tabular}{c c c c}
        \hline
        \hline
          & Grid & Genetic A & Genetic B\\
        \hline
        \hline
        Population size & - & 5 & 8 \\
        Max generations & - & 10 & 5 \\
        Crossover & - & 0.5 & 0.5\\
        Mutation & - & 0.2  & 0.4 \\
       \hline
       \hline
         &   Hyperparameter results &   \\
       \hline
        Layers & 4 & 4  & 4\\
      Nodes & 200 & 200 & 100\\
      Learning rate & 0.001 & 0.001 & 0.001\\
      Batch size & 64 & 16  & 16\\
      \hline
      \hline
         &   Metrics  &  &\\
       \hline
      MSE & 0.0356 & 0.0154 & 0.0193\\
      MAE & 0.1074 & 0.0583 & 0.0571\\
      $R^2$ & 0.7057 & 0.8608 & 0.8192\\
      \hline
      \hline
      Evaluations & 32 & 31 & 25\\
      
       \hline
       \hline
    \end{tabular}
    \caption{\changes{Results of ANN training for photometric redshift}}
    \label{tab:results_photoz}
\end{table}

For this case, the neural network model has $6$ nodes in the input layer (the attributes of Table \ref{tab:photoz}) and 1 node in the output layer, corresponding to the photometric redshift. Using the hyperparameter grid varying the number of layers, number of nodes, batch size and learning rate (with the values shown in Table \ref{tab:hyp_all}), we have a total of 32 combinations. And we choose three different configurations for the genetic algorithms regarding to the crossover and mutation probabilities, population size and generations.

Table \ref{tab:results_photoz} contains the results of the hyperparameter tuning with the grid and different configurations of genetic algorithms (cases A and B). We found that the genetic algorithms A and B, considering the values of the MSE, MAE and $R^2$, are better than the grid method. In both cases, the MSE and MAE are around half of the values obtained by the grid; and the $R^2$ score is also significantly higher.

\begin{figure*}[t!]
        \captionsetup{justification=raggedright, singlelinecheck=false, font=footnotesize}
        \centering
        \makebox[14cm][c]{
        \includegraphics[trim= 0mm 0mm 10mm 0mm, clip, width=7cm, height=7cm]{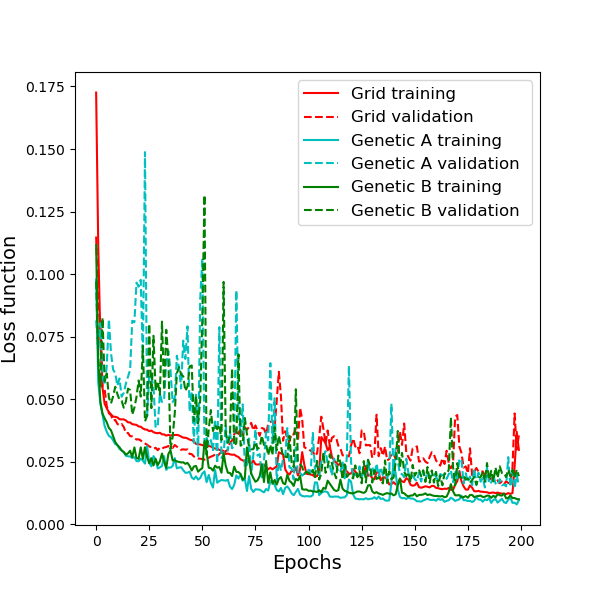}
        \includegraphics[trim= 0mm 0mm 5mm 0mm, clip, width=7.5cm, height=7cm]{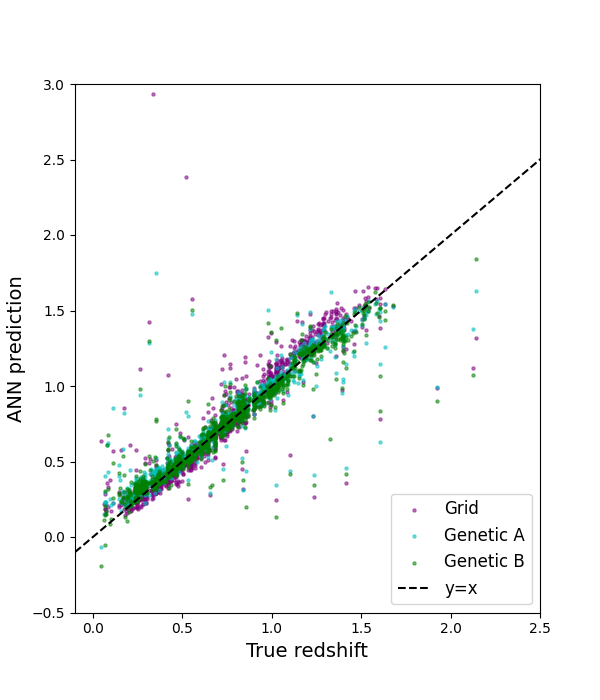}

        }
        \caption{\textit{Left }: Behavior of the loss functions. \textit{Right}: Predicted photometric redshifths by the ANNs in comparison with the true spectroscopic redshifts }
        \label{fig:results_photoz}
\end{figure*}

During the training, in the two cases, the genetic algorithms have a lower value in the loss function; we plot them in Figure \ref{fig:results_photoz}.  The second panel of Figure \ref{fig:results_photoz} shows the predicted redshift versus the true redshift and the three results are visually similar than the presented in Reference \cite{zhou2019deep}, and the predictions of the neural networks proposed by the genetic algorithms have less dispersion than that corresponding to the grid method.

\section{Discussion}
\label{sec:conclusions}

Throughout the three cases in this work, we conclude that the use of the genetic algorithms is an interesting strategy to find a correct neural network architecture in a cosmological context. 

It is noteworthy that when genetic algorithms use higher mutation probability values, fewer neural network architectures are evaluated, more variance is induced in individuals between subsequent generations and this may allow a good combination of hyperparameters to be found more quickly. We can notice the effect of the mutation probability in the number of evaluations, smaller mutation probability leads to  more neural network evaluations. However, we must remember that higher values of the mutation probability can become a random walk and should be avoided.

We have observed that the hyperparameter grid evaluates each combination just once, regardless of its proximity to the optimal value. In contrast, the genetic algorithms evaluate more times the configurations that are within a region of the search space where the optimum is likely to be, they may even evaluate the same point more than once; for example the mutation or crossover of two different individuals could generate the same new point. This behavior makes the solution found by the genetic algorithms more reliable because the individuals of the last population have been tested better than the configurations evaluated within the hyperparameter grid framework. In general, as can be noticed in the more extensive exploration of Table \ref{tab:ap_jla_tests}, the genetic algorithms found better solutions than the grid method.

We tested the genetic algorithms with a relatively small population size and number of generations and, in the majority of the analyzed cases, their performance were very competitive and even better than the traditional grid; \changes{we validate this observation with the complementary analysis in the Appendix \ref{appendixA}}. In this paper we have tested the hyperparameter tuning with genetic algorithms only with feedforward neural networks and in three very specific examples; nevertheless, this same methodology can be used in other cosmological applications and with any other types of neural networks. Moreover, in cosmological scenarios where more numerical precision, more complex architectures or larger search space size (i.e., more number of hyperparameters) are required, then the genetic algorithms are still expected to perform very well.

Another remark is that the process of running a genetic algorithm to find the hyperparameters of a neural network can be slow, if we increase the initial population and decrease the mutation probability, it can be similar to the hyperparameter grid method (see Table \ref{tab:ap_longgenetic}). However, in the current times of precision cosmology we believe it is a necessary cost to obtain better neural network models for the observational evidence.

\section*{Acknowledgements}

JAV acknowledges the support provided by FOSEC SEP-CONACYT Investigaci\'on B\'asica A1-S-21925, Ciencias de Frontera CONACYT-PRONACES/304001/202 and UNAM-DGAPA-PAPIIT IA104221. IGV thanks the CONACYT postdoctoral fellowship, the support of the ICF-UNAM, Ricardo Medel Esquivel for the discussions on genetic algorithms and Viviana Acquaviva for her workshop on machine learning at the VIII Essential Cosmology for the Next Generation.

\appendix

\section{\changes{A note about the randomness}}
\label{appendixA}
The reader may have noticed that the performance of the genetic algorithms, in the above three examples, is always better than the hyperparameter grid. A valid question may be why; if the grid considers all combinations of a proposed set of hyperparameters and hence the individuals considered in the genetic algorithms are a subset of the grid.
The cause of this apparent problem is the randomness of all the processes involved; even fixing the random seed, the values of the weights in the neural networks and the value of the mean square error at the end of training are very similar, but not exactly the same due to their random nature and the way computers generate the pseudo-random numbers. On the other hand, the hyperparameter grid only evaluates each combination once, while the genetic algorithm can evaluate the same individual several times over generations, so the genetic algorithm obtains the best solution even considering the small fluctuations related to the randomness of the process.

In order to validate the previous statement, we have performed 10 times the example of the Section \ref{sec:case1_rec}. Genetic algorithms A, B and C refer to the values of the Table \ref{tab:results_jla}, in which all of them have a crossover probability of 0.5 and a mutation probability of 0.5, 0.2 and 0.4, respectively. Results are shown in Table \ref{tab:ap_jla_tests}. In general, we can notice that the performance of the genetic algorithms is better than the grid method and with less variance in their metrics. 

A remarkable fact is the decrease in the number of evaluations for neural networks architectures. However, because the randomness, in a few cases the grid method can be better than the genetic algorithm. This issue can be tackled with more generations and a higher population in the genetic algorithm; for this purpose, see Table \ref{tab:ap_longgenetic}, the test uses the configuration of the Genetic Algorithm B for the Section \ref{sec:case1_rec} but with a population size of 15 and 10 as maximum number of generations. We can notice that the standard deviation decreases and in all ten runs the genetic algorithm is better than the grid method, shown in Table \ref{tab:ap_jla_tests}, and yet this genetic algorithm executes less evaluation of neural network architectures.

\begin{table}[h!]
     \small
    \centering
    \begin{tabular}{c c c c c c c c c}
    \hline
    \# Test	&	Layers	&	Nodes	&	Learning rate	&	Batch size	&	Evaluations	&	MSE	&	MAE	&	$R^2$	\\
    \hline
    \hline
	&  &   &   & Grid &    &   &   & \\												
    \hline																			
        1	&	4	&	200	&	0.00010	&	16	&	128	&	0.0371	&	0.1165	&	0.4968	\\
        2	&	2	&	50	&	0.00100	&	16	&	128	&	0.0370	&	0.1141	&	0.6914	\\
        3	&	4	&	150	&	0.00010	&	16	&	128	&	0.0361	&	0.1133	&	0.5394	\\
        4	&	3	&	200	&	0.00100	&	16	&	128	&	0.0350	&	0.1093	&	0.4285	\\
        5	&	4	&	100	&	0.00100	&	16	&	128	&	0.0448	&	0.1222	&	0.4994	\\
    	6	&	3	&	200	&	0.00100	&	16	&	128	&	0.0807	&	0.1714	&	0.4531	\\
    	7	&	4	&	150	&	0.00010	&	16	&	128	&	0.0409	&	0.1189	&	0.6572	\\
    	8	&	3	&	200	&	0.00100	&	16	&	128	&	0.0678	&	0.1478	&	0.6657	\\
    	9	&	3	&	200	&	0.00100	&	16	&	128	&	0.0369	&	0.1101	&	0.4174	\\
    	10	&	4	&	100	&	0.00100	&	16	&	128	&	0.0340	&	0.1035	&	0.6679	\\
\hline																			
        Average	$\pm$ std dev &	&		&	&	&	128	$\pm$ 0.0 &	0.0450	$\pm$ 0.016 &	0.1227	$\pm$ 0.021 &	0.5517	$\pm$ 0.109 \\
    \hline
    \hline
	&  &   &   & Genetic A &    &   &   & \\												
    \hline																			
    	1	&	2	&	100	&	0.00010	&	2	&	41	&	0.0311	&	0.0998	&	0.6735	\\
    	2	&	4	&	50	&	0.00010	&	2	&	41	&	0.0363	&	0.1069	&	0.5894 \\
    	3	&	2	&	200	&	0.00010	&	2	&	36	&	0.0315	&	0.0996	&	0.6893	\\
    	4	&	2	&	100	&	0.00010	&	2	&	36	&	0.0313	&	0.0997	&	0.6866	\\
        5	&	4	&	150	&	0.00010	&	4	&	39	&	0.0340	&	0.1049	&	0.7091\\
    	6	&	3	&	100	&	0.00010	&	2	&	33	&	0.0363	&	0.1070	&	0.6357\\
    	7	&	2	&	100	&	0.00010	&	2	&	38	&	0.0316	&	0.0998	&	0.6515\\
    	8	&	1	&	50 &	0.00100	&	2	&	37	&	0.0323	&	0.1023	&	0.4249\\
    	9	&	2	&	200	&	0.00010	&	2	&	37	&	0.0316	&	0.0997	&	0.6787\\
    	10	&	2	&	200	&	0.00100	&	8	&	31	&	0.0411	&	0.1157	&	0.6956	\\
\hline																			
     Average	$\pm$ std dev	&	& 	& & & 37	$\pm$ 3	& 0.0337 $\pm$0.003	 &	0.104 $\pm$	 0.005 & 0.6434 $\pm$ 0.084 \\
    \hline
    \hline
	&  &   &   & Genetic B &    &   &   & \\												
    \hline																				
    	1	&	2	&	200	&	0.00010	&	2	&	34	&	0.0314	&	0.0985	&	0.6797	\\
    	2	&	4	&	200	&	0.00010	&	2	&	36	&	0.0470	&	0.1285	&	0.7054	\\
    	3	&	4	&	150	&	0.00010	&	2	&	36	&	0.0407	&	0.1177	&	0.6659	\\
    	4	&	3	&	100	&	0.00010	&	4	&	32	&	0.0333	&	0.1009	&	0.6912	\\
        5	&	4	&	100	&	0.00100	&	4	&	31	&	0.0515	&	0.1449	&	0.0920	\\
    	6	&	4	&	100	&	0.00010	&	4	&	28	&	0.0342	&	0.1075	&	0.6933	\\
    	7	&	3	&	100	&	0.00010	&	4	&	35	&	0.0343	&	0.1029	&	0.6792	\\
    	8	&	3	&	200	&	0.00100	&	16	&	31	&	0.0386	&	0.1102	&	0.6795	\\
    	9	&	4	&	50	&	0.00010	&	4	&	29	&	0.0333	&	0.1021	&	0.7204	\\
    	10	&	2	&	100	&	0.00010	&	2	&	22	&	0.0305	&	0.0982	&	0.6779	\\
\hline																			
     Average	$\pm$ std dev	&	&		&	&	&	31 $\pm$4	&	0.0375$\pm$	0.007	&	0.1111$\pm$0.015 &	0.6285$\pm$	0.189\\
    \hline
    \hline
	&  &   &   & Genetic C &    &   &   & \\												
    \hline																				
    	1	&	4	&	100	&	0.00010	&	4	&	50	&	0.0336	&	0.1059	&	0.7251	\\
    	2	&	1	&	200	&	0.00100	&	4	&	38	&	0.0339	&	0.1043	&	0.3435	\\
    	3	&	4	&	50	&	0.00010	&	4	&	51	&	0.0330	&	0.1021	&	0.4566	\\
    	4	&	2	&	100	&	0.00010	&	2	&	48	&	0.0314	&	0.0980	&	0.6970	\\
        5	&	2	&	200	&	0.00100	&	8	&	52	&	0.0357	&	0.1120	&	0.5074	\\
    	6	&	1	&	200	&	0.00100	&	4	&	55	&	0.0319	&	0.1015	&	0.3628	\\
    	7	&	4	&	100	&	0.00010	&	4	&	52	&	0.0328	&	0.1039	&	0.6673	\\
    	8	&	2	&	150	&	0.00010	&	2	&	48	&	0.0317	&	0.0989	&	0.6083	\\
    	9	&	3	&	100	&	0.00010	&	4	&	40	&	0.0329	&	0.1022	&	0.6479	\\
    	10	&	2	&	100	&	0.00010	&	2	&	49	&	0.0338	&	0.1006	&	0.6747	\\
\hline																			
	Average $\pm$ std dev	&	&	&	&	&	48 $\pm$5 	&	0.0331	$\pm$ 0.001	&	0.1029	$\pm$ 0.004	&	0.5691$\pm$ 0.141\\
\hline																			
    \end{tabular}
    \caption{10 tests for each method using the JLA SNe-Ia dataset.}
    \label{tab:ap_jla_tests}
\end{table}

\begin{table}[t!]
    \small
    \centering
    \begin{tabular}{c c c c c c c c c}
    \hline
    $\#$ Run	&	Layers	&	Nodes	&	Learning rate	&	Batch size	&	Evaluations	&	MSE	&	MAE	&	$R^2$\\
    \hline
    \hline
    1	&	3	&	200	&	0.0001	&	4	&	108	&	0.0313	&	0.0996	&	0.6414	\\
    2	&	4	&	50	&	0.0001	&	4	&	110	&	0.0320	&	0.0997	&	0.6893	\\
    3	&	1	&	100	&	0.001	&	4	&	106	&	0.0331	&	0.1016	&	0.606	\\
    4	&	1	&	100	&	0.001	&	4	&	95	&	0.0328	&	0.1028	&	0.4797	\\
    5	&	3	&	100	&	0.0001	&	4	&	112	&	0.0330	&	0.1027	&	0.6823	\\
    6	&	1	&	100	&	0.001	&	4	&	92	&	0.0322	&	0.1021	&	0.4873	\\
    7	&	3	&	100	&	0.0001	&	4	&	99	&	0.0318	&	0.1000	&	0.6242	\\
    8	&	3	&	100	&	0.0001	&	4	&	103	&	0.0338	&	0.1024	&	0.6372	\\
    9	&	1	&	100	&	0.001	&	4	&	83	&	0.0347	&	0.1027	&	0.6965	\\
    10	&	4	&	100	&	0.0001	&	4	&	92	&	0.0326	&	0.1047	&	0.6929	\\
    \hline
    Average $\pm$ std dev		&	&	&	&	&	100 $\pm$ 9	&	0.0327	$\pm$0.001 	&	0.1018$\pm$0.002	&0.6237	$\pm$ 0.080\\
    \hline
    \end{tabular}
    \caption{Tests for a genetic algorithm with 15 individuals as population, mutation 0.2, crossover 0.5 and with 10 maximum number of generations.}
    \label{tab:ap_longgenetic}
\end{table}

\newpage
\bibliographystyle{unsrt}
\bibliography{references.bib}

\end{document}